\documentclass[a4paper,11pt]{article}
\pdfoutput=1 % if your are submitting a pdflatex (i.e. if you have              
\usepackage{jcappub} % for details on the use of the package, please            
\usepackage[T1]{fontenc} % if needed       

\usepackage{array}

\title{\boldmath Conservative Constraints on Early Cosmology: an illustration of the {\sc Monte Python} cosmological parameter inference code} 

\author[a]{Benjamin Audren,}
\author[a,b,c]{Julien Lesgourgues,}
\author[d]{Karim Benabed,}
\author[d]{and Simon Prunet}

\affiliation[a]{Institut de Th\'eorie des Ph\'enom\`enes Physiques,\\ \'Ecole Polytechnique F\'ed\'erale de Lausanne,\\ CH-1015, Lausanne, Switzerland}
\affiliation[b]{CERN, Theory Division,\\ CH-1211 Geneva 23, Switzerland}
\affiliation[c]{LAPTh (CNRS -Universit\'e de Savoie), BP 110,\\ F-74941 Annecy-le-Vieux Cedex, France}
\affiliation[d]{Institute d'Astrophysique de Paris, 98 Bd Arago, F-75014 Paris, France}

\emailAdd{benjamin.audren@epfl.ch}
\emailAdd{Julien.Lesgourgues@cern.ch}
\emailAdd{benabed@iap.fr}
\emailAdd{prunet@iap.fr}

\abstract{
Models for the latest stages of the cosmological evolution rely on a less solid
theoretical and observational ground than the description of earlier stages
like BBN and recombination. As suggested in a previous work by Vonlanthen et
al., it is possible to tweak the analysis of CMB data in such way to avoid
making assumptions on the late evolution, and obtain robust constraints on
``early cosmology parameters''. We extend this method in order to marginalise
the results over CMB lensing contamination, and present updated results based
on recent CMB data. Our constraints on the minimal early cosmology model are
weaker than in a standard $\Lambda$CDM analysis, but do not conflict with this
model. Besides, we obtain conservative bounds on the effective neutrino number
and neutrino mass, showing no hints for extra relativistic degrees of freedom,
and proving in a robust way that neutrinos experienced their non-relativistic
transition after the time of photon decoupling. This analysis is also an
occasion to describe the main features of the new parameter inference code {\sc
Monte Python}, that we release together with this paper. {\sc Monte Python} is
a user-friendly alternative to other public codes like {\sc CosmoMC},
interfaced with the Boltzmann code {\sc class}.}

\begin{document}
\maketitle
\flushbottom

\section{Introduction}

Models for the evolution of the early universe between a redshift of a few
millions and a few hundreds have shown to be very predictive and successful:
the self-consistency of Big Bang Nucleosynthesis (BBN) model could be tested by
comparing the abundance of light elements and the result of Cosmic Microwave
Background (CMB) observations concerning the composition of the early universe;
the shape of CMB acoustic peaks matches accurately the prediction of
cosmological perturbation theory in a Friedmann-Lema\^{\i}tre Universe
described by general relativity, with a thermal history described by standard
recombination. The late cosmological evolution is more problematic. Models for
the acceleration of the universe, based on a cosmological constant, or a dark
energy component, or departures from general relativity, or finally departure
from the Friedmann-Lema\^{\i}tre model at late times, have shown no predictive
power so far. The late thermal history, featuring reionization from stars, is
difficult to test with precision. Overall, it is fair to say that ``late
cosmology'' relies on less solid theoretical or observational ground than
``early cosmology''.

When fitting the spectrum of temperature and polarisation CMB anisotropies, we
make simultaneously some assumptions on early and late cosmology, and obtain
intricate constraints on the two stages. However, Vonlanthen et
al.~\cite{Vonlanthen:2010cd} suggested a way to carry the analysis leading to
constraints only on the early cosmology part. This is certainly interesting
since such an analysis leads to more robust and model-indepent bounds than a
traditional analysis affected by priors on the stages which are most poorly
understood. The approach of \cite{Vonlanthen:2010cd}  avoids making assumptions
on most relevant ``late cosmology-related effects'': projection effects due to
the background evolution, photon rescattering during reionization, and the late
Integrated Sachs Wolfe (ISW) effect.

In this work, we carry a similar analysis, pushed to a higher precision level
since we also avoid making assumptions on the contamination of primary CMB
anisotropies by weak lensing. We use the most recent available data from the
Wilkinson Microwave Anisotropy Probe (WMAP) and South Pole Telescope (SPT)
data, and consider the case of a minimal ``early cosmology'' model, as well as
extended models with free density of ultra-relativistic relics or massive
neutrinos.

This analysis is an occasion to present a new cosmological parameter inference
code. This Monte Carlo code written in Python, called {\sc Monte
Python}\footnote{\tt http://montepython.net}, offers a convenient alternative
to {\sc CosmoMC} \cite{Lewis:2002ah}. It is interfaced with the Boltzmann code
{\sc class}\footnote{\tt http://class-code.net}
\cite{Lesgourgues:2011re,Blas:2011rf}.  {\sc Monte Python} is released publicly
together with this work.

In section 2, we explain the  method allowing to get constraints only on the
early cosmological evolution. We present our result for the minimal early
cosmology model in section 3, and for two extended models in section 4. In
section 5, we briefly summarize some of the advantages of {\sc Monte Pyhton},
without entering into technical details (presented anyway in the code
documentation). Our conclusions are highlighted in section 6.

\section{How to test early cosmology only?}

\begin{figure}[tbp]
\centering
\includegraphics[height=7cm]{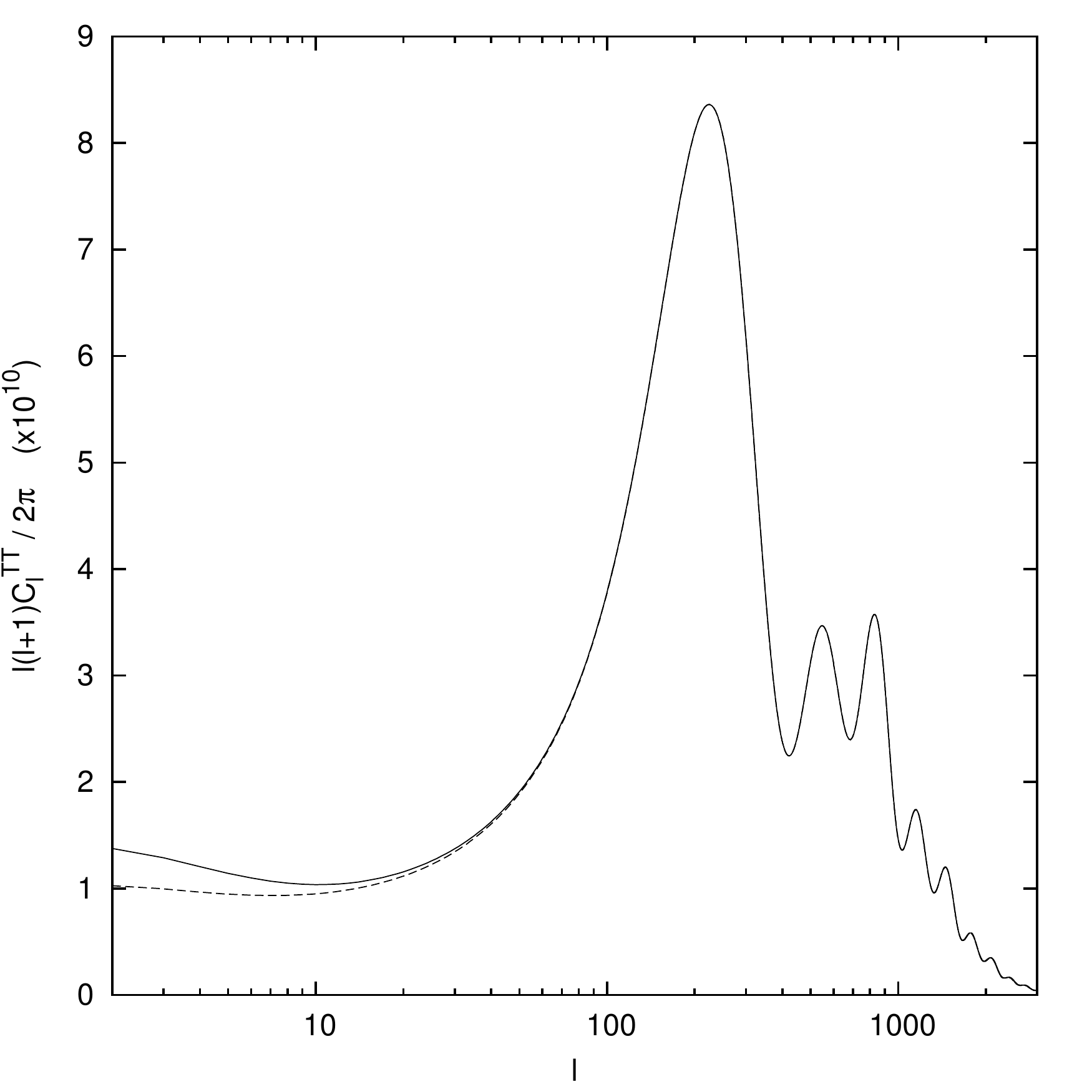}
\includegraphics[height=7cm]{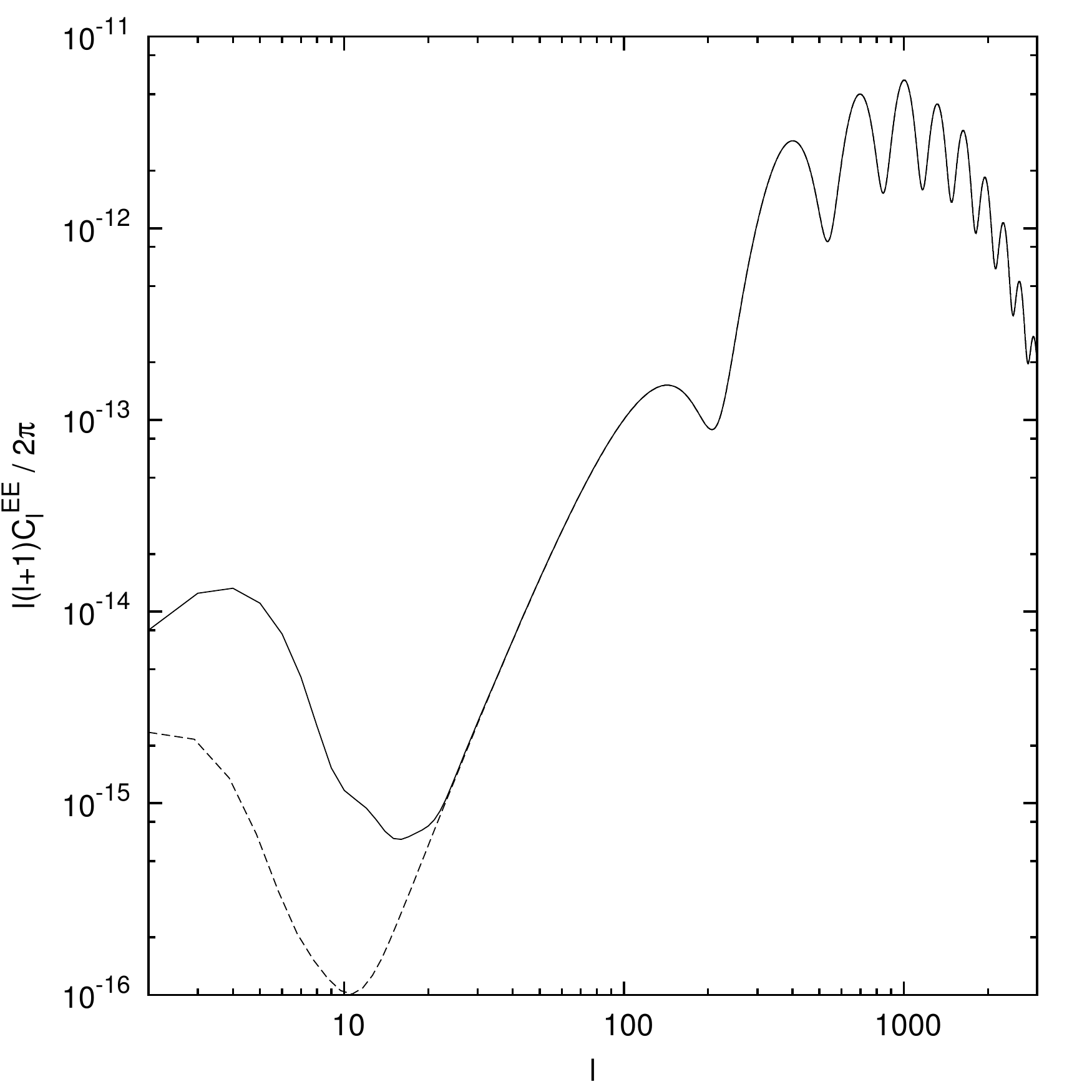}
\caption{Dimensionless temperature (left) and E-polarization (right) unlensed spectra of two
$\Lambda$CDM models with the same value of ``early cosmology'' parameters
($\omega_b$, $\omega_{cdm}$, $A_s$, $n_s$) (fixed to WMAP best-fitting values),
and different values of ``late cosmology'' parameters: ($\Omega_\Lambda$,
$z_{reio}$) = (0.720,10) (solid curves) or (0.619, 5) (dashed curve). The
dashed curves have been rescaled vertically by the ratio of $e^{-2 \tau}$ and
horizontally by the ratio of  $d_A(\tau_{rec})$ in each model, using the values
of $\tau$ and $d_A(\tau_{rec})$ calculated by {\sc class} for each model. At
$\ell=40$, the difference between the dashed and solid line in the temperature
plot is under $2\,\mu$K$^2$. At $\ell=80$, it is already below $1\,\mu$K$^2$.
\label{fig:rescale}}
\end{figure}

The spectrum of primary CMB temperature anisotropies is sensitive to various
physical effects:
\begin{itemize}
\item (C1) the location  of the acoustic peaks in multipole space depends on
  the sound horizon at decoupling $d_s(\tau_{rec})$ (an ``early
  cosmology''-dependent parameter) divided by the angular diameter distance to
  decoupling $d_A(\tau_{rec})$ (a ``late cosmology''-dependent parameter,
  sensitive to the recent background evolution: acceleration, spatial
  curvature, etc.)
\item (C2) the contrast between odd and even peaks depends on
  $\omega_b/\omega_\gamma$, i.e. on ``early cosmology''.
\item (C3) the amplitude of all peaks further depends on the amount of
  expansion between radiation-to-matter equality and decoupling, governing the amount of
  perturbation damping at the beginning of matter domination, and on the amount
  of early integrated Sachs-Wolfe effect enhancing the first peak just after
  decoupling. These are again ``early cosmology'' effects (in the minimal
  $\Lambda$CDM model, they are both regulated by the redshift
  of radiation-to-matter equality, i.e by $\omega_m/\omega_r$). 
\item (C4) the enveloppe of high-$\ell$ peaks depends on the diffusion damping
  scale at decoupling $\lambda_d(\tau_{rec})$ (an  ``early cosmology''
  parameter) divided again by the angular diameter distance to decoupling
  $d_A(\tau_{rec})$ (a ``late cosmology'' parameter).
\item (C5-C6) the global shape depends on initial conditions through the
  primordial spectrum amplitude $A_s$ (C5) and tilt $n_s$ (C6), which are both
  ``early cosmology'' parameters.
\item (C7) the slope of the temperature spectrum at low $\ell$ is affected by the
  late integrated Sachs Wolfe effect, i.e. by ``late cosmology''. This effect
  could actually be considered as a contamination of the primary spectrum by
  secondary anisotropies, which are not being discussed in this list.
\item (C8) the global amplitude of the spectrum at $\ell\gg40$ is reduced by the
  late reionization of the universe, another ``late cosmology'' effect. The
  amplitude of this suppression is given by $e^{-2 \tau}$, where $\tau$ is the
  reionization optical depth. 
\end{itemize}
In summary, primary CMB temperature anisotropies are affected by late cosmology
only through: (i) projection effects from real space to harmonic space,
controlled by $d_A(\tau_{rec})$; (ii) the late ISW effect, affecting only small
$\ell$'s; and (iii) reionization, suppressing equally all multipoles at $\ell\gg40$ .
These are actually the sectors of the cosmological model which are most poorly
constrained and understood. But we see that the shape of the power spectrum at
$\ell\gg40$, interpreted modulo an arbitrary scaling in amplitude ($C_{\ell}
\rightarrow \alpha C_{\ell}$) and in position ($C_{\ell} \rightarrow C_{\beta \ell}$),
contains information on early cosmology only. This statement is very general
and valid for extended cosmological models. In the case of the $\Lambda$CDM
models, it is illustrated by figure~\ref{fig:rescale}, in which we took two
different $\Lambda$CDM models (with different late-time geometry and
reionization history), and rescaled one of them with a shift in amplitude given
by $e^{-2 \tau-\tau'}$ and in scale given by $d_A/d_A'$. At $\ell\gg40$, the two
spectra are identical. For more complicated cosmological models sharing the
same physical evolution until approximately $z\sim100$, a similar rescaling and
matching would work equally well.

If polarization is taken into account, the same statement remains valid. The
late time evolution affects the polarization spectrum through the  angular
diameter distance to decoupling $d_A(\tau_{rec})$ and through the impact of
reionization, which also suppresses the global amplitude at $\ell\gg40$, and
generates an additional feature at low $\ell$'s, due to photon re-scatering by
the inhomogeneous and ionized inter-galactic medium. The shape of the primary
temperature and polarization spectrum at $\ell\gg40$, interpreted modulo a global
scaling in amplitude and in position, only contains information on the early
cosmology.

However, the CMB spectrum that we observe today gets a contribution from
secondary anisotropies and foregrounds. In particular, the observed CMB spectra
are significantly affected by CMB lensing caused by large scale structures.
This effect depends on the small scale matter power spectrum, and therefore on
late cosmology (acceleration, curvature, neutrinos becoming non-relativistic at
late time, possible dark energy perturbations, possible departures from
Einstein gravity on very large scales, etc.). In the work of
\cite{Vonlanthen:2010cd}, this effect was mentioned but not dealt with, because
of the limited precision of WMAP5 and ACBAR data compared to the amplitude of
lensing effects, at least within the multipole range studied in that paper ($40
\leq \ell \leq 800$). The results that we will present later confirm that this
simplification was sufficient and did not introduce a significant ``late
cosmology bias''. However, with the full WMAP7+SPT data (that we wish to use up
to the high multipoles), it is not possible to ignore lensing, and in order to
probe only early cosmology, we are forced to marginalize over the lensing
contamination, in the sense of the method described below. By doing so, we will
effectively get rid of the major two sources of secondary (CMB)
anisotropies, the late ISW effect and CMB lensing. We neglect the impact of
other secondary effects like the Rees-Sciama effect. As far as foregrounds are
concerned, the approach of WMAP and SPT consists in eliminating them with a
spectral analysis, apart from residual foregrounds which can be fitted to the
data, using some nuisance parameters which are marginalized over. By following
this approach, we also avoid to introduce a ``late cosmology bias'' at the
level of foregrounds.

Let us now discuss how one can marginalize over lensing corrections.  Ideally,
we should lens the primary CMB spectrum with all possible lensing patterns, and
marginalize over the parameters describing these patterns.  But the lensing of
the CMB depends on the lensing potential spectrum $C_{\ell}^{\phi \phi}$, that can
be inferred from the matter power spectrum at small redshift, $P(k,z)$. We
should marginalize over all possible shapes for $C_{\ell}^{\phi \phi}$, i.e. over an
infinity of degrees of freedom. We need to find a simpler approach.

One can start by noticing that modifications of the late-time background
evolution caused by a cosmological constant, a spatial curvature, or even some
inhomogeneous cosmology models, tend to affect matter density fluctuations in a
democratic way: all Fourier modes being inside the Hubble radius and on linear
scales are multiplied by the same redshift-dependent growth factor. CMB lensing
is precisely caused by such modes. Hence, for this category of models,
differences in the late-time background evolution lead to a different amplitude
for $C_{\ell}^{\phi \phi}$, and also a small tilt since different $\ell$'s probe the
matter power spectrum at different redshifts. Hence, if we fit the temperature
and polarization spectrum at $\ell\gg40$ modulo a global scaling in amplitude, a
global  shift in position, and additionally an arbitrary scaling and tilting of
the lensing spectrum $C_{\ell}^{\phi \phi}$ that one would infer assuming
$\Lambda$CDM, we still avoid making assumption about the late-time evolution.

There are also models introducing a scale-dependent growth factor, i.e.
distortions in the shape of the matter power spectrum. This is the case in
presence of massive neutrinos or another hot dark matter component, of dark
energy with unusually large perturbations contributing to the total perturbed
energy-momentum tensor, or in modified gravity models. In principle, these
effects could lead to arbitrary distortions of $C_{\ell}^{\phi \phi}$ as a function
of $\ell$. Fortunately, CMB lensing only depends on the matter power spectrum
$P(k,z)$ integrated over a small range of redshifts and wave numbers. Hence it
makes sense to stick to an expansion scheme:
at first order we can account for the effects of a scale-dependent growth factor
by writing the power spectrum as the one predicted by $\Lambda$CDM cosmology,
multiplied by arbitrary rescaling and tilting factors; and at the next order,
one should introduce a running of the tilt, then a running of the running, etc.
By marginalizing over the rescaling factor, tilting factor, running, etc., one
can still fit the CMB spectra without making explicit assumptions about the
late-time cosmology. In the result section, we will check that the information
on early cosmology parameters varies very little when we omit to marginalize
over the lensing amplitude, or when we include this effect, or when we also
marginalize over a tilting factor. Hence we will not push the analysis to the
level of an arbitrary lensing running factor.

\section{Results assuming a minimal early cosmology model}

We assume a ``minimal early cosmology'' model described by four parameters
($\omega_b$, $\omega_{cdm}$, $A_s$, $n_s$).  In order to extract constraints
independent of the late cosmological evolution, we need to fit the CMB
temperature/polarisation spectrum measured by WMAP (seven year data
\cite{Komatsu:2010fb}) and SPT \cite{Reichardt:2011yv} only above a given value
of $\ell$ (typically $\ell \sim 40$), and to marginalize over two factors accounting
for vertical and a horizontal scaling. In practice, there are several ways in
which this could be implemented.

For the amplitude, we could fix the reionization history and simply marginalize
over the amplitude parameter $A_s$. By fitting the data at $\ell \gg 40$, we
actually constrain the product $e^{-2 \tau_{reio}} A_s$, i.e. the primordial
amplitude rescaled by the reionization optical depth $\tau_{reio}$,
independently of the details of  reionization. In our runs, we fix
$\tau_{reio}$ to an arbitrary value, and we vary $A_s$; but in the Markov
chains, we keep memory of the value of the derived parameter $e^{-2 \tau_{reio}}
A_s$. By quoting bounds on $e^{-2 \tau} A_s$ rather than $A_s$, we avoid making
explicit assumptions concerning the reionization history.

For the horizontal scaling, we could modify {\sc class} in such way to use
directly $d_A(\tau_{rec})$ as an input parameter. For input values of
($\omega_b$, $\omega_{cdm}$, $d_A(\tau_{rec})$), {\sc class} could in principle
find the correct spectrum at $\ell \gg 40$. It is however much simpler to use the
unmodified code and pass values of the five parameters ($\omega_b$,
$\omega_{cdm}$, $A_s$, $n_s$, $h$). In our case, $h$ should not be interpreted
as the reduced Hubble rate, but simply as a parameter controlling the value of
the physical quantity $d_A(\tau_{rec})$. For any given set of parameters, the
code computes the value that $d_A(\tau_{rec})$ would take in a $\Lambda$CDM
model with the same early cosmology and with a Hubble rate $H_0=100h$km/s/Mpc.
It then fits the theoretical spectrum to the data. The resulting likelihood
should be associated to the inferred value of $d_A(\tau_{rec})$ rather than to
$h$. The only difference between this simplified approach and that in which
$d_A(\tau_{rec})$ would be passed as an input parameter is that in one case,
one assumes a flat prior on $d_A(\tau_{rec})$, and in the other case a flat
prior on $h$. But given that the data allows $d_A(\tau_{rec})$ to vary only
within a very small range where it is almost a linear function of $h$, the
prior difference has a negligible impact.

To summarize, in order to get constraints on ``minimal early cosmology'', it is
sufficient to run Markov Chains in the same way as for a minimal $\Lambda$CDM
model with parameters ($\omega_b$, $\omega_{cdm}$, $A_s$, $n_s$, $\tau$, $h$),
excepted that:
\begin{itemize}
\item we do not fit the lowest temperature/polarization multipoles to the data;
\item we fix $\tau$ to an arbitrary value;
\item we do not plot nor interpret the posterior probability of the parameters
  $A_s$ and $h$. We only pay attention to the posterior probability of the two
  derived parameters $e^{-2 \tau} A_s$ and $d_A(\tau_{rec})$, which play the
  role of the vertical and horizontal scaling factors, and which are
  marginalized over when quoting bounds on the remaining three ``early
  cosmology parameters'' ($\omega_b$, $\omega_{cdm}$, $n_s$).
\end{itemize}
Hence, for a parameter inference code, this is just a trivial matter of
defining and storing two ``derived parameters''. For clarity, we will refer to
the runs performed in this way as the ``agnostic'' runs.

\begin{table}[tbp]
\centering
{\small
\setlength{\extrarowheight}{6pt}
\begin{tabular}{|c|ccccccc|}
\hline
& $100~\omega_{b }$  & $\omega_{cdm }$  & $n_{s }$  & ${d_A}^{rec}$  & $10^{9}e^{-2 \tau} A_{s }$ & $A_{lp}$ & $n_{lp}$ \\ \hline
%&&&&&&&\\
&\multicolumn{7}{c|}{$\Lambda$CDM}\\
%&&&&&&&\\
& 
$2.241_{-0.044}^{+0.043}$ & $0.1114_{-0.0048}^{+0.0048}$ & $0.960_{-0.011}^{+0.011}$ & $12.93_{-0.12}^{+0.11}$ & $2.069_{-0.092}^{+0.085}$
&& \\ 
&&&&&&&\\
\hline
%&&&&&&&\\
&\multicolumn{7}{c|}{same lensing potential as in $\Lambda$CDM}\\
%&&&&&&&\\
$\ell\geq40$&
$2.204_{-0.047}^{+0.048}$ & $0.1160_{-0.0059}^{+0.0056}$ & $0.946_{-0.014}^{+0.014}$ & $12.85_{-0.13}^{+0.13}$ & $2.20_{-0.13}^{+0.12}$&&\\
$\ell\geq60$&
$2.203_{-0.053}^{+0.050}$ & $0.1163_{-0.0065}^{+0.0063}$ & $0.945_{-0.016}^{+0.016}$ & $12.84_{-0.14}^{+0.14}$ & $2.20_{-0.15}^{+0.13}$&&\\
$\ell\geq80$&
$2.190_{-0.057}^{+0.053}$ & $0.1180_{-0.0073}^{+0.0067}$ & $0.940_{-0.018}^{+0.019}$ & $12.81_{-0.15}^{+0.15}$ & $2.26_{-0.18}^{+0.15}$&&\\
$\ell\geq100$&
$2.184_{-0.056}^{+0.054}$ & $0.1187_{-0.0079}^{+0.0067}$ & $0.935_{-0.019}^{+0.020}$ & $12.80_{-0.15}^{+0.16}$ & $2.29_{-0.20}^{+0.16}$&& \\
&&&&&&&\\\hline
%&&&&&&&\\
&\multicolumn{7}{c|}{marginalization over lensing potential amplitude}\\
%&&&&&&&\\
$\ell\geq100$&
$2.159_{-0.064}^{+0.060}$ & $0.1227_{-0.0088}^{+0.0083}$ & $0.926_{-0.022}^{+0.022}$ &  $12.73_{-0.17}^{+0.18}$ & $2.39_{-0.23}^{+0.20}$ & $0.88_{-0.13}^{+0.12}$&\\
&&&&&&&\\\hline
%&&&&&&&\\
&\multicolumn{7}{c|}{marginalization over lensing potential amplitude and tilt}\\
%&&&&&&&\\
$\ell\geq100$&
$2.160_{-0.068}^{+0.064}$ & $0.1222_{-0.0094}^{+0.0088}$ & $0.927_{-0.024}^{+0.024}$ & $12.74_{-0.18}^{+0.18}$ & $2.38_{-0.25}^{+0.20}$ & $0.78_{-0.15}^{+0.20}$ & $-0.16_{-0.33}^{+0.55}$\\
&&&&&&&\\
\hline
\end{tabular}
}
\caption{Limits at the $68\%$ confidence level of the mininum credible interval
of model parameters.  The $\Lambda$CDM model of the first line has a sixth
independent parameter ($z_{reio}$) that we do not show. We do not show either
the limits on the three nuisance parameters associated to the SPT
likelihood.\label{tab:minimal}}
\end{table}

\begin{figure}[tbp]
\centering
\includegraphics[width=12cm]{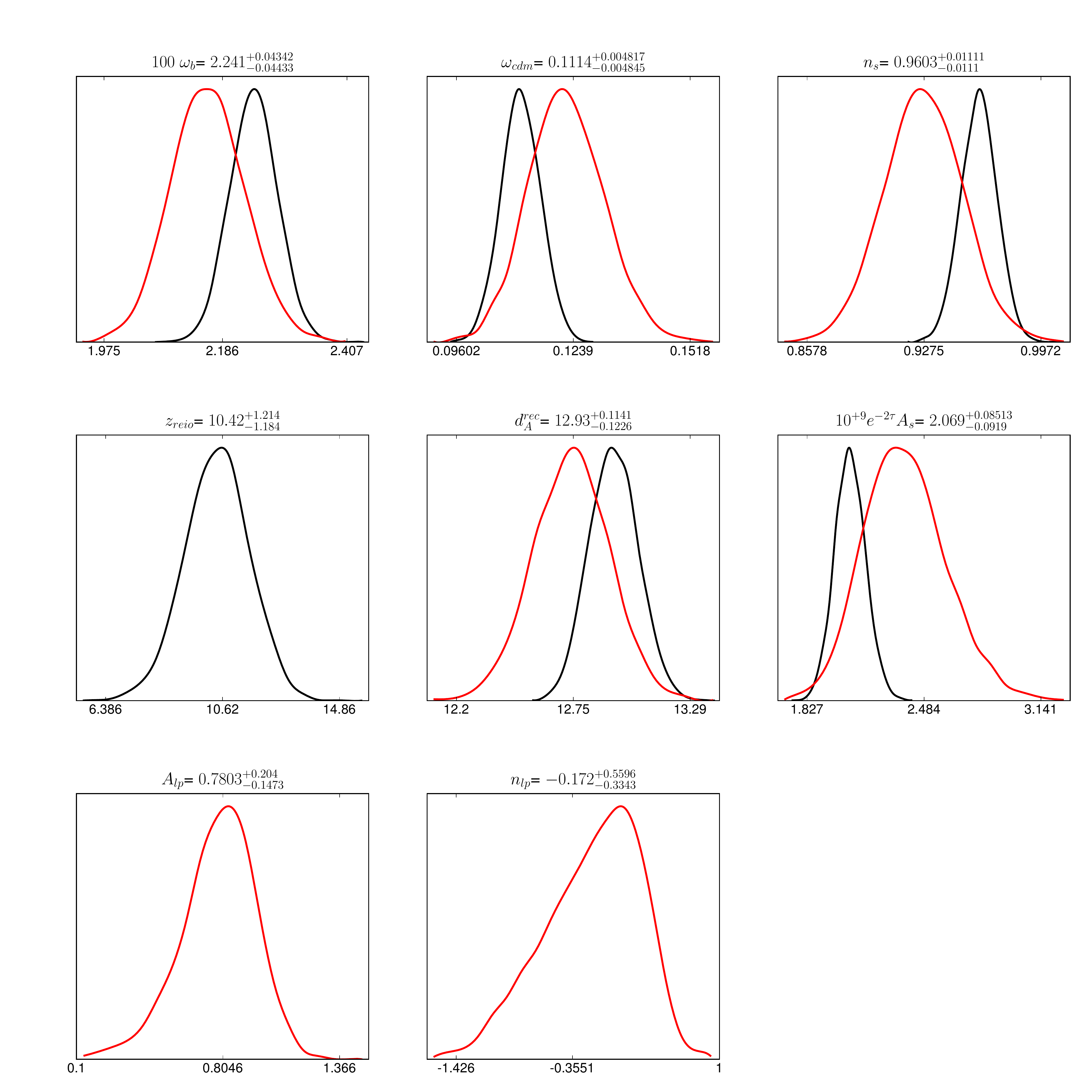}
\caption{Constraints on the five parameters of the minimal early cosmology model
(red), compared to usual constraints on the minimal $\Lambda$CDM model (black).
The $\Lambda$CDM has a sixth independent parameter, the reionization optical
depth. The constraints on early cosmology (called ``agnostic constraints'' in
the text) includes a marginalization over the amplitude and tilt of the matter
power spectrum leading to CMB lensing. We do not show here the posterior of the
three nuisance parameter used to fit SPT data.\label{fig:1d}}
\end{figure}

In the second line of Table~\ref{tab:minimal}, we show the bounds obtained with
such an agnostic run, for a cut-off value $\ell=40$. These results can be compared
with those of a minimal $\Lambda$CDM model, obtained through the same machinery
but with all multipoles $\ell\geq 2$. Since the agnostic bounds rely on less
theoretical assumptions, they are slightly wider. Interestingly, the central
value of $\omega_b$ and $n_s$ are smaller in absence of late-cosmology priors,
and larger for $\omega_{cdm}$. Still the $\Lambda$CDM results are compatible
with the agnostic results, which means that on the basis of this test, we
cannot say that $\Lambda$CDM is a bad model. Our agnostic bounds on
($\omega_b$, $\omega_{cdm}$, $n_s$) are simply more model-independent and
robust, and one could argue that when using CMB bounds in the study of BBN, in
CDM relic density calculations or for inflationary model building, one should
better use those bounds in order to avoid relying on the most uncertain
assumptions of the minimal cosmological model, namely $\Lambda$ domination and
standard reionization.

The decision to cut the likelihood at $\ell\geq 40$ was somewhat arbitrary.
Figure~\ref{fig:rescale} shows that two rescaled temperature spectra with
different late-time cosmology tend only gradually towards each other above
$\ell\sim 40$. We should remove enough low multipoles in order to be sure that
late time cosmology has a negligible impact given the data error bars. We
tested this dependence by cutting the likelihood at $\ell\geq 60$, $\ell\geq 80$ or
$\ell\geq 100$. When increasing the cut-off from 40 to 100, we observe variations
in the mean value that are less important than from 2 to 40. To have the more
robust constraints, we will then take systematically the cut-off of $\ell=100$,
which is the one more likely to avoid any contamination from ``late time
cosmology''.

%We observe a very small variation of the bounds when increasing
%the cut-off from 40 to 60, an even smaller one when going from 60 to 80, and
%then the results remain stable between 80 and 100. We conclude that our results
%are robust against the choice of a particular cut-off value, and for simplicity
%we perform all the remaining runs with a cut-off at $\ell=100$.

Until now, our analysis is not completely ``agnostic'', because we did not
marginalize over lensing. We fitted the data with a lensed power spectrum,
relying on the same lensing potential as an equivalent $\Lambda$CDM model with
the same values of ($\omega_b$, $\omega_{cdm}$, $n_s$, $e^{-2 \tau} A_s$,
$d_A(\tau_{rec})$). To deal with lensing, we introduce three new parameters
($A_{lp}$, $n_{lp}$, $k_{lp}$) in {\sc class}. Given the traditional input
parameters ($\omega_b$, $\omega_{cdm}$, $A_s$, $n_s$, $h$), the code first
computes the Newtonian potential $\phi(k,z)$. This potential is then rescaled
as
\begin{equation}
\phi(k,z) \longrightarrow A_{lp} \left(\frac{k}{k_{lp}}\right)^{n_{lp}}\phi(k,z)~.
\end{equation}
Hence, the choice ($A_{lp}$, $n_{lp}$)=(1,0) corresponds to the standard
lensing potential predicted in the $\Lambda$CDM model. Different values
correspond to an arbitrary rescaling or tilting of the lensing potential, which
can be propagated consistently to the lensed CMB temperature/polarization
spectrum.

The sixth run shown in Table~\ref{tab:minimal} corresponds to $n_{lp}=0$ and a
free parameter $A_{lp}$. The minimum credible interval for this rescaling
parameter is $A_{lp}=0.88^{+0.12}_{-0.13}$ at the 68\% Confidence Level (CL), and is
compatible with one. This shows that WMAP7+SPT data alone are sensitive to
lensing, and well compatible with the lensing signal predicted by the minimal
$\Lambda$CDM model. It is also interesting to note that the bounds on other
cosmological parameter move a little bit, but only by a small amount (compared
to the difference between the $\Lambda$CDM and the previous ``agnostic'' runs),
showing that ``agnostic bounds'' are robust.

In the seventh line of Table~\ref{tab:minimal}, we also marginalize over the
tilting parameter $n_{lp}$ (with unbounded flat prior). A priori, this
introduces a lot of freedom in the model. Nicely, this parameter is still well
constrained by the data ($n_{lp}=-0.16^{+0.55}_{-0.33}$ at 68\%CL), and compatible
with the $\Lambda$CDM prediction $n_{lp}=0$. Bounds on other parameters vary
this time by a completely negligible amount: this motivates us to stop the
expansion at the level of $n_{lp}$, and not to test the impact of  running. The
credible interval for $A_{lp}$ is the only one varying significantly when
$n_{lp}$ is left free, but this result depends on the pivot scale $k_{lp}$,
that we choose to be equal to $k_{lp}=0.1$/Mpc, so that the amplitude of the
lensing spectrum $C_{\ell}^{\phi \phi}$ is nearly fixed at $\ell\sim 100$. By tuning
the pivot scale, we could have obtained bounds on $A_s$ nearly equal for the
case with/without free $n_{lp}$. The posterior probability of each parameter
marginalized over other parameters is shown in Figure~\ref{fig:1d}, and
compared with the results of the  standard $\Lambda$CDM analysis.

Our results nicely agree with those of \cite{Vonlanthen:2010cd}. These authors
found a more pronounced drift of the parameters ($\omega_b$, $\omega_{cdm}$,
$n_s$) with the cut-off multipole than in the first part of our analysis, but
this is because we use data on a wider multipole range and have a larger lever
arm. Indeed, Ref.~ \cite{Vonlanthen:2010cd} limited their analysis of WMAP5
plus ACBAR data to $\ell\leq800$, arguing that above this value, lensing would
start playing an important role. In our analysis, we include WMAP7 plus 47 SPT
band powers probing up to $\ell \sim 3000$, but for consistency we must
simultaneously marginalize over lensing. Indeed, the results of Ref.~
\cite{Vonlanthen:2010cd} are closer to our results with lensing marginalization
(the fully ``agnostic'' ones) that without. Keeping only one digit in the error
bar, we find ($100\omega_b=2.16\pm0.07$, $\omega_{cdm} = 0.122\pm0.009$,
$n_s=0.93\pm0.02$), when this reference found ($100\omega_b=2.13\pm0.05$,
$\omega_{cdm} = 0.124\pm0.007$, $n_s=0.93\pm0.02$). The two sets of results are
very close to each other, but our central values for $\omega_b$ and
$\omega_{cdm}$ are slightly closer to the $\Lambda$CDM one. The fact that we
get slightly larger error bars in spite of using better data in a wider
multipole range is related to our lensing marginalization: we see that by
fixing lensing, this previous analysis was implicitly affected by a partial
``late cosmology prior'', but only at a very small level. 

Our results from the last run can be seen as robust ``agnostic'' bounds on
($\omega_b$, $\omega_{cdm}$, $n_s$), only based on the  ``minimal early
cosmology'' assumption. They are approximately twice less constraining than
ordinary $\Lambda$CDM models, and should be used in conservative studies of the
physics of BBN, CDM decoupling and inflation.

\section{Effective neutrino number and neutrino mass}

We can try to generalize our analysis to extended cosmological models. It would
make no sense to look at models with spatial curvature, varying dark energy or
late departures from Einstein gravity, since all these assumptions would alter
only the late time evolution, and our method is designed precisely in such way
that the results would remain identical. However, we can explore models with
less trivial assumptions concerning the early cosmological evolution. This
includes for instance models with:
\begin{itemize}
\item
a free primordial helium fraction $Y_{He}$. So far, we assumed $Y_{He}$ to be a
function of $\omega_b$, as predicted by standard BBN (this is implemented in
{\sc class} following the lines of Ref.~\cite{Hamann:2007sb}). Promoting
$Y_{He}$ as a free parameter would be equivalent to relax the assumption of
standard BBN. Given the relatively small sensitivity of current CMB data to
$Y_{He}$ \cite{Komatsu:2010fb}, we do not perform such an analysis here, but
this could be done in the future using e.g. Planck data.
\item
a free density of relativistic species, parametrized by a free effective
neutrino number $N_{\rm eff}$, differing from its value of  $3.046$ in the
minimal $\Lambda$CDM model~\cite{Mangano:2001iu}. This parameter affects the
time of equality between matter and radiation, but this effect can be cancelled
at least at the level of ``early cosmology'' by tuning appropriately the
density of barons and CDM. Even in that case, relativistic species will leave a
signature on the CMB spectrum, first through a change in the diffusion damping
scale $\lambda_d(\tau_{rec})$, and second  through direct effects at the level
of perturbations, since they induce a gravitational damping and phase shifting
of the photon fluctuation \cite{Hu:1995en,Bashinsky:2003tk}. It is not obvious
to anticipate up to which level these effects are degenerate with those of
other parameters. Hence it is interesting to run Markov chains and search for
``agnostic bounds'' on $N_{\rm eff}$.
\item
neutrino masses (or for simplicity, three degenerate masses $m_\nu$ summing up
to $M_\nu=3m_\nu$).  Here we are not interested in the fact that massive
neutrinos affect the background evolution and change the ratio between the
redshift of radiation-to-matter equality, and that of matter-to-$\Lambda$
equality. This is a ``late cosmology'' effect that we cannot probe with our
method, since we are not sensitive to the second equality. However, for masses
of the order of $m_\nu\sim0.60$~eV, neutrinos become non-relativistic at the
time of photon decoupling. Even below this value, the mass leaves a signature
on the CMB spectrum coming from the fact that, first, they are not yet
ultra-relativistic at decoupling, and second, the transition to the
non-relativistic regime takes place when the CMB is still probing metric
perturbations through the early integrated Sachs-Wolfe effect. Published bounds
on $M_\nu$ from CMB data alone probe all these intricate effects
\cite{Komatsu:2010fb}, and it would be instructive to obtain robust bounds
based only on the mass impact on ``early cosmology''.
\end{itemize}

\begin{table}[tbp]
\centering
{\small
\setlength{\extrarowheight}{6pt}
\begin{tabular}{|cccccccc|}
\hline
$100~\omega_{b }$  & $\omega_{cdm }$  & $n_{s }$  & $N_{\rm eff}$ & ${d_A}^{rec}$  & $10^{9}e^{-2 \tau} A_{s }$ & $A_{lp}$ & $n_{lp}$ \\ \hline
%&&&&&&&\\
\multicolumn{8}{|c|}{$\Lambda$CDM}\\
%&&&&&&&\\
$2.279_{-0.056}^{+0.053}$ & $0.124_{-0.013}^{+0.011}$ & $0.979_{-0.019}^{+0.019}$ & $3.77_{-0.66}^{+0.58}$ &$12.35_{-0.53}^{+0.46}$ & $2.01_{-0.10}^{+0.10}$
&& \\ 
&&&&&&&\\
\hline
%&&&&&&&\\
\multicolumn{8}{|c|}{$\ell \geq 100$, marginalization over lensing potential amplitude and tilt}\\
%&&&&&&&\\

$2.03_{-0.16}^{+0.13}$ &
$0.113_{-0.015}^{+0.011}$ & $0.862_{-0.077}^{+0.065}$ &
$2.04_{-1.26}^{+0.78}$ & $13.59_{-0.87}^{+0.96}$ &
$2.84_{-0.59}^{+0.49}$ &
$0.69_{-0.18}^{+0.21}$ &
$-0.23_{-0.43}^{+0.57}$\\
&&&&&&&\\
\hline
\end{tabular}
}
\caption{Limits at the 68\% confidence level of the minimum credible interval
of model parameters.  The $\Lambda$CDM+$N_{\rm eff}$ model of the first line has
a seventh independent parameter ($z_{reio}$) that we do not show. We do not
show either the limits on the three nuisance parameters associated to the SPT
likelihood. \label{tab:neff}}
\end{table}

\begin{figure}[tbp]
\centering
\includegraphics[width=15cm]{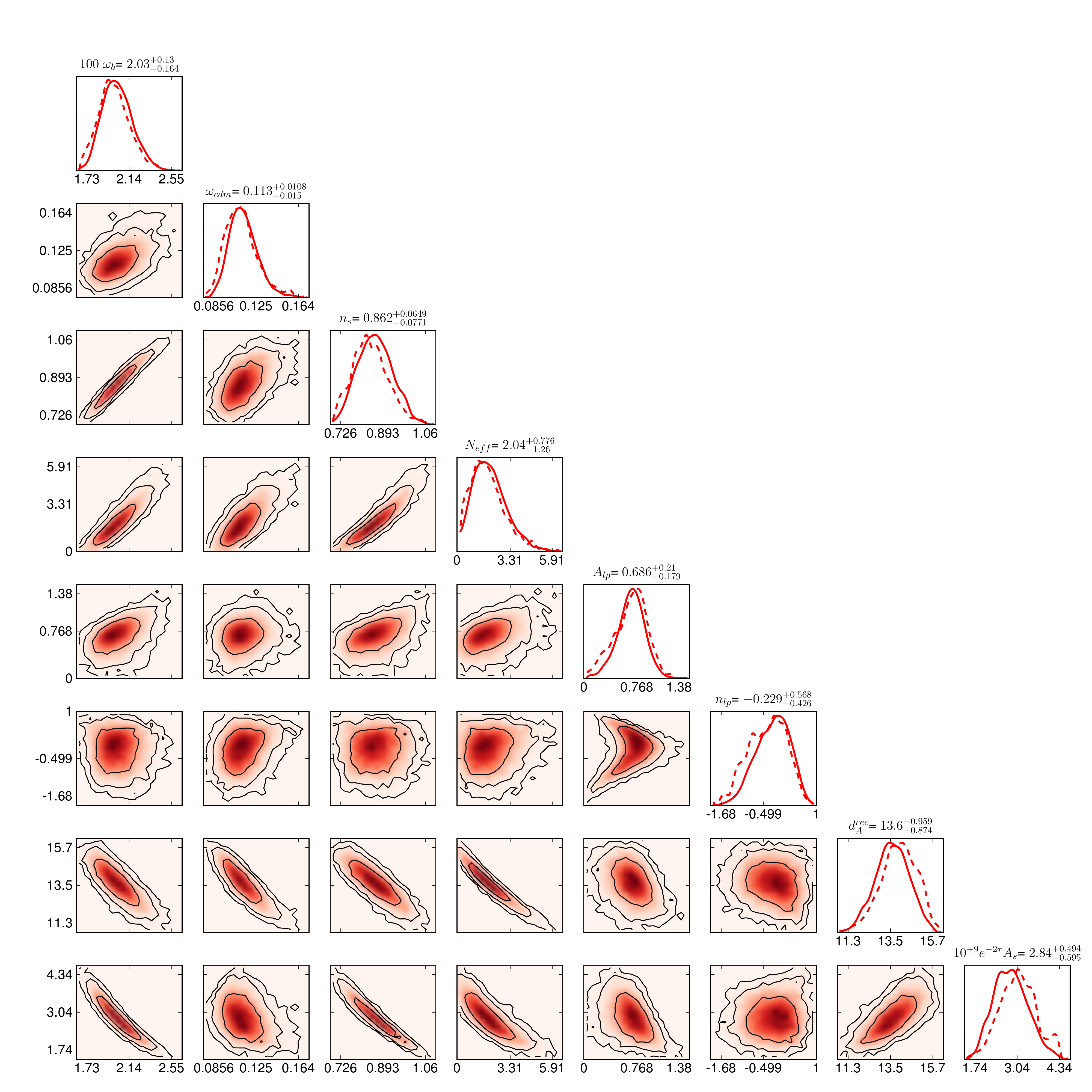}
\caption{One and two-dimensional posterior distribution (solid line) of
the parameters of the ``agnostic'' run with a free effective neutrino number.
The dashed line stands for the average likelihood distribution. The concentric
contour lines in the two-dimensional posteriors stand for 68, 95 and 99\% CL.
\label{fig:triangle-neff}}
\end{figure}

For the effective neutrino number, we performed two runs similar to our
previous $\Lambda$CDM and ``fully agnostic'' run (with marginalization over
lensing amplitude and tilt), in presence of one additional free parameter
$N_{\rm eff}$. Our results are summarized in Table~\ref{tab:neff} and
Figure~\ref{fig:triangle-neff}. In the $\Lambda$CDM+$N_{\rm eff}$ case, we get
$N_{\rm eff}=3.77^{+0.58}_{-0.66}$ (68\% CL), very close to the result of
\cite{Keisler:2011aw}, $N_{\rm eff}=3.85\pm0.62$ (differences in the priors can
explain this insignificant difference). It is well-known by now that the
combination of WMAP and small-scale CMB data shows a marginal preference for
extra relativistic degrees of freedom in the seven-parameter model. The
surprise comes from our ``agnostic'' bound on this number, $N_{\rm
eff}=2.04^{+0.78}_{-1.26}$ (68\% CL). As explained before, this bound cannot come
from a change in the time of equality, nor in the scale of the first peak,  nor
in the late integrated Sachs-Wolfe effect; it can only result from the
measurement of the the sound horizon $d_s(\tau_{rec})$ {\it relatively} to the
diffusion damping scale $\lambda_d(\tau_{rec})$, and from the direct effects of
extra relativistic degrees of freedom on photon perturbations. Hence it is
normal that $N_{\rm eff}$ is much less constrained in the agnostic runs, but
the interesting conclusion is that without assuming $\Lambda$CDM at late time,
the CMB does not favor high values of $N_{\rm eff}$. It is compatible with the
standard value $N_{\rm eff}=3.046$ roughly at the one-$\sigma$ level, with even
a marginal preference for smaller values. This shows that recent hints for
extra relativistic relics in the universe disappear completely if we discard
any information on the late time cosmological evolution. It is well-known that
$N_{\rm eff}$ is very correlated with $H_0$ and affected by the inclusion of
late cosmology data sets, like direct measurement of $H_0$ or of the BAO scale.
Our new result shows that even at the level of CMB data only, the marginal hint
for large $N_{\rm eff}$ is driven by physical effects related to late cosmology
(and in particular by the angular diameter distance to last scattering as
predicted in $\Lambda$CDM).

The triangle plot in Figure~\ref{fig:triangle-neff} shows that in the agnostic
run, $N_{\rm eff}$ is still very correlated with other parameters such as
$\omega_b$, $\omega_{cdm}$ and $n_s$. Low values of $N_{\rm eff}$
(significantly smaller than the standard value 3.046) are only compatible with
a very small $\omega_b$, $\omega_{cdm}$ and $n_s$. Note that in this work, we
assume standard BBN in order to predict $Y_{He}$ as a function of $\omega_b$
(and of $N_{\rm eff}$ when this parameter is also left free), but we do not
incorporate data on light element abundances. By doing so, we would favor the
highest values of $\omega_b$ in the range allowed by the current analysis
($\omega_b \sim 0.022$), and because of parameter correlations we would also
favor the highest values of $\omega_{cdm}$, $n_s$ and $N_{\rm eff}$, getting
close to the best-fitting values in the minimal early cosmology model with
$N_{\rm eff}\sim3.046$.

\begin{table}[tbp] \centering {\small \setlength{\extrarowheight}{6pt}
  \begin{tabular}{|cccccccc|} \hline $100~\omega_{b }$  & $\omega_{cdm }$  &
    $n_{s }$  & $M_{\nu}$ (eV) & ${d_A}^{rec}$  & $10^{9}e^{-2 \tau} A_{s }$ &
    $A_{lp}$ & $n_{lp}$ \\ \hline
%&&&&&&&\\
\multicolumn{8}{|c|}{$\Lambda$CDM}\\
%&&&&&&&\\
$2.205_{-0.049}^{+0.046}$ & $0.114_{-0.0050}^{+0.0052}$ &
$0.949_{-0.013}^{+0.014}$ & $<1.4$ & $12.86_{-0.13}^{+0.13}$ &
$2.16_{-0.12}^{+0.10}$ && \\ &&&&&&&\\ \hline
%&&&&&&&\\
\multicolumn{8}{|c|}{$\ell \geq 100$, marginalization over lensing potential
amplitude and tilt}\\
%&&&&&&&\\
$2.136_{-0.072}^{+0.065}$ & $0.123_{-0.010}^{+0.009}$ &
$0.920_{-0.025}^{+0.025}$ & $<1.8$ & $12.69_{-0.19}^{+0.19}$ &
$2.43_{-0.28}^{+0.21}$ & $0.81_{-0.16}^{+0.22}$ & $-0.11_{-0.32}^{+0.58}$ \\
%&& \\
&&&&&&&\\ \hline \end{tabular} }\ \caption{Limits at the 68\% confidence level
of the minimum credible interval of model parameters (excepted for $M_\nu$, for
which we show the 95\% CL upper limit).  The $\Lambda$CDM+$M_{\nu}$ model of
the first line has a seventh independent parameter ($z_{reio}$) that we do not
show. We do not show either the limits on the three nuisance parameters
associated to the SPT likelihood.  \label{tab:mnu}} \end{table}

\begin{figure}[tbp]
\centering
\includegraphics[width=15cm]{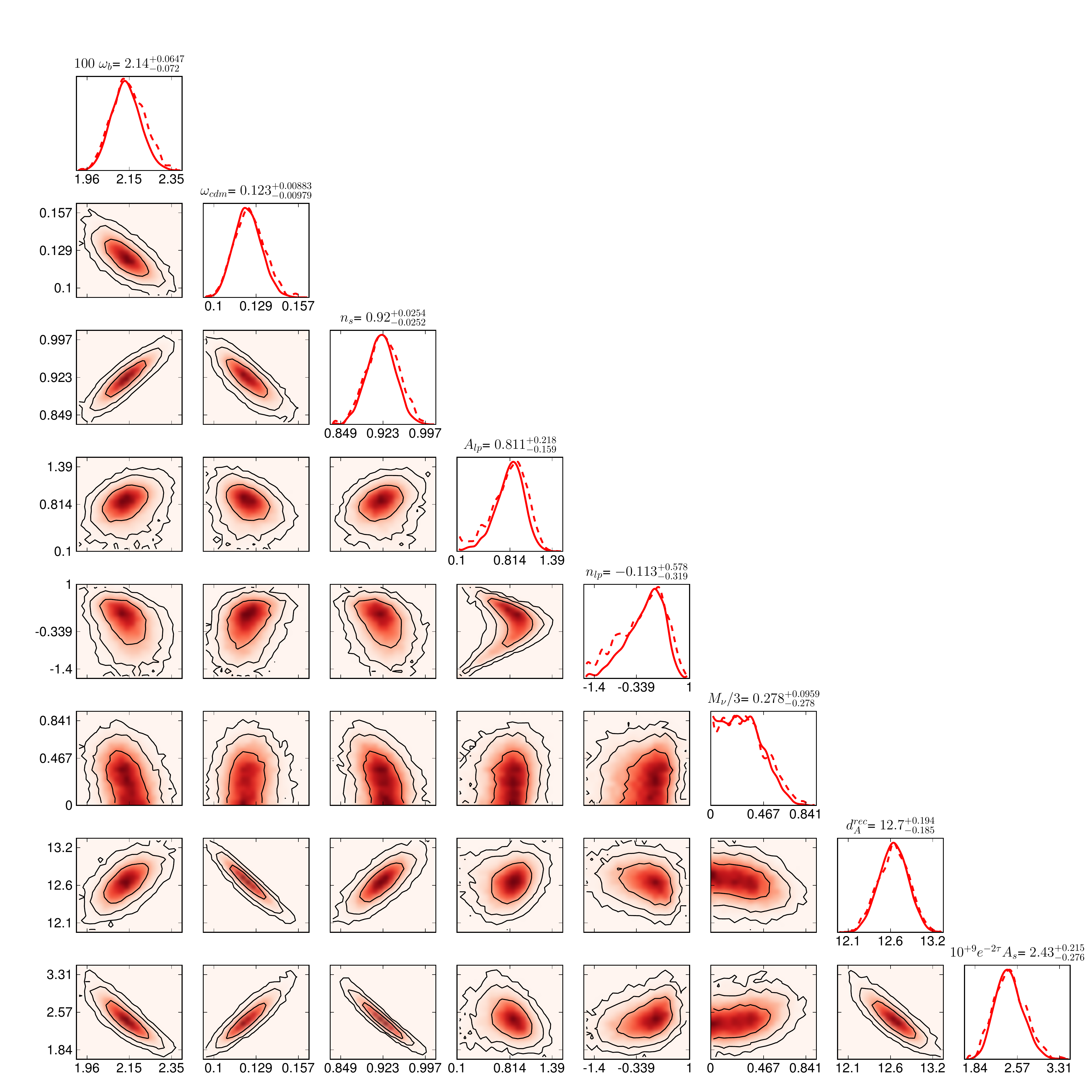}
\caption{One and two-dimensional posterior distribution (solid line) of
the parameters of the ``agnostic'' run with a total neutrino mass $M_\nu$
(assuming three degenerate neutrinos of individual mass $m_\nu$). Again, dashed
line stands for average likelihood distribution, contour lines indicate the 68,
95 and 99\% CL.}\label{fig:triangle-mnu}
\end{figure}

For neutrino masses, we performed two similar runs (summarised in
Table~\ref{tab:mnu} and Figure~\ref{fig:triangle-mnu}), with now $M_\nu$
being the additional parameter (assuming three degenerate neutrino species). In
the $\Lambda$CDM case, our result $M_\nu<1.4$~eV (95\%CL) is consistent with
the rest of the literature, and close to the WMAP-only bound of
\cite{Komatsu:2010fb}: measuring the CMB damping tail does not bring
significant additional information on the neutrino mass. In the agnostic run,
this constraint only degrades to $M_\nu<1.8$~eV (95\%CL). This limit is
consistent with the idea that for sufficiently large $m_\nu$, the CMB can set a
limit on the neutrino mass not just through its impact on the background
evolution at late time (and its contribution to $\omega_m$ today), but also
through direct effects occurring at the time of recombination and soon after.
It is remarkable that this is true even for neutrinos of individual mass
$m_\nu\sim0.6$~eV, becoming non-relativistic precisely at the time of photon
decoupling. The conclusion that the CMB is not compatible with neutrinos
becoming non-relativistic before $z_{rec}$ (and not even slightly before!)
appears to be very robust, and independent of any constraint on the late
cosmological evolution.

\section{Advantages of {\sc Monte Python}}

The results of this paper were obtained with the new parameter inference code
{\sc Monte Python}, that we release publicly together with this article.
Currently, {\sc Monte Python} is interfaced with the Boltzmann code {\sc
class}, and explores parameter space with the Metropolis-Hastings algorithm,
just like {\sc CosmoMC}\footnote{In this paper we refer to the version of
CosmoMC available at the time of submitting, i.e. the version of October 2012.}
(note however that interfacing it with other codes and switching to other
exploration algorithms would be easy, thanks to the modular architecture of the
code). Hence, the difference with {\sc CosmoMC} \cite{Lewis:2002ah} does not
reside in a radically different strategy, but in several details aiming at
making the user's life easy. It is not our goal to describe here all the
features implemented in {\sc Monte Python}: for that, we refer the reader to
the documentation distributed with the code. We only present here a brief
summary of the main specificities of {\sc Monte Python}.

\medskip

{\bf Language and compilation.} As suggested by its name, {\sc Monte Python} is
a Monte Carlo code written in Python. This high-level language allows to code
with a very concise style: {\sc Monte Python} is compact, and the
implementation of e.g. new likelihoods requires very few lines. Python is also
ideal for wrapping other codes from different languages: {\sc Monte Python} needs
to call {\sc class}, written in C, and the WMAP likelihood code, written in
Fortran 90. The user not familiar with Python should not worry: for most
purposes, {\sc Monte Python} does not need to be edited, when {\sc CosmoMC}
would need to: this is explained in the fourth paragraph below.\\ 
Another advantage of Python is that it includes many libraries (and an easy way
to add more), so that {\sc Monte Python} is self-contained. Only the WMAP
likelihood code needs its own external libraries, as usual. Python codes do not
require a compilation step. Hence, provided that the user has Python 2.7
installed on his/her computer\footnote{The documentation explains how to run
with Python 2.6. The code would require very minimal modifications to run with
Python 3.0.} alongside very standard modules, the code only needs to be
downloaded, and is ready to work with.

\medskip

{\bf Modularity.} A parameter inference code is based on distinct blocks: a
likelihood exploration algorithm, an interface with a code computing
theoretical predictions (in our case, a Boltzmann code solving the cosmological
background and perturbation evolution), and an interface with each experimental
likelihood. In {\sc Monte Python}, all three blocks are clearly split in
distinct modules. This would make it easy, e.g., to interface {\sc Monte
Python} with {\sc camb} \cite{Lewis:1999bs} instead of {\sc class}, or to
switch from the in-build Metropolis-Hastings algorithm to another method, e.g.
a nested sampling algorithm.\\
The design choice of the code has been to write these modules as different
classes, in the sense of C++, whenever it served a purpose. For instance, all
likelihoods are defined as separated classes. It allows for easy and intuitive
way of comparing two runs, and helps simplify the code. The cosmological module
is also defined as a class, with a defined amount of functions. If someone
writes a python wrapper for {\sc camb} defining these same functions, then {\sc
Monte Python} would be ready to serve.\\
On the other hand the likelihood exploration part is contained in a normal
file, defining only functions. The actual computation is only done in the file
{\tt code/mcmc.py}, so it is easy to implement a different exploration
algorithm. From the rest of the code, this step would be as transparent as
possible.\\
In Python, like in C++, a class can inherit properties from a parent class.
This becomes particularly powerful when dealing with data likelihoods. Each
likelihood will inherit from a basic {\tt likelihood} class, able to read data
files, and to treat storage. In order to implement a new likelihood, one then
only needs to write the computation part, leaving the rest automatically done
by the main code. This avoids several repetitions of the same piece of code.
Furthermore, if the likelihood falls in a generic category, like CMB
likelihoods based on reading a file in the format {\tt .newdat} (same files as
in {\sc CosmoMC}), it will inherit more precise properties from the {\tt
likelihood\_newdat} class, which is itself a daughter of the {\tt likelihood}
class. Hence, in order to incorporate CMB likelihoods apart from WMAP, one only
needs to write one line of python for each new case: it is enough to tell,
e.g., to the class accounting for the CMB experiment SPT that it inherits all
properties from the generic {\tt likelihood\_newdat} class. Then, this class is
ready to read a file in the  {\tt .newdat} format and to work. Note that our
code already incorporates another generic likelihood class that will be useful
in the future for reading Planck likelihoods, after the release of Planck
data.\\
Finally, please note that these few lines of code to write for a new likelihood
are completely outside the main code containing the exploration algorithm, and
the cosmological module. You do not need to tell the rest of the code that you
wrote something new, you just have to use your new likelihood by its name in a
starting parameter file.

\medskip

{\bf Memory keeping and safe running.} Each given run, i.e. each given
combination of a set of parameters to vary, a set of likelihoods to fit, and a
version of the Boltzmann code, is associated to a given directory where the
chains are written (e.g. it could be a directory called {\tt
chains/wmap\_spt/lcdm}). All information about the run is logged automatically
in this directory, in a file {\tt log.param}, at the time when the first chain
is started. This file contains the parameter names, ranges and priors, the list
of extra parameters, the version of the Boltzmann code, the version and the
characteristics of each data likelihood, etc. Hence the user will always
remember the details of a previous run. \\
Moreover, when a new chain is started, the code reads this log file (taking
full advantage of the class structure of the code). If the user started the new
chain with an input file, the code will compare all the data in the input file
with the data in the {\tt log.param} file. If they are different, the code
complains and stop. The user can then take two decisions: either some
characteristic of the run has been changed without noticing, and the input file
can be corrected. Or it has been changed on purpose, then this is a new run and
the user must require a different output directory. This avoids the classical
mistake of mixing unwillingly some chains that should not be compared to each
other. Now, if the input file is similar to the {\tt log.param} file, the chain
will start (it will not take the same name as previous chains: it will append
automatically to its name a number equal to the first available number in the
chain directory). In addition, the user who wishes to launch new chains for the
same run can omit to pass an input file: in this case all the information about
the run is automatically read in the {\tt log.param} and the chain can start.\\
The existence of  {\tt log.param} file has another advantage. When one wants to
analyze chains and produce result files and plots (the equivalent of running
{\tt Getdist} and matlab or maple in the case of {\sc CosmoMC}), one simply needs
to tell {\sc Monte Python} to analyze a given directory. It is not needed to pass
another input file, since all information on parameter names and ranges will be
found in the {\tt log.param}. If the output needs to be customized (i.e.,
changing the name of the parameters, plotting only a few of them, rescaling
them by some factor, etc.), then the user can use command lines and eventually
pass one small input file with extra information.

\medskip

{\bf No need to edit the code when adding parameters.} The name of cosmological
parameters is never defined in {\sc Monte Python}. The code only knows that in
the input file, it will read a list of parameter names (e.g. {\tt omega\_b},
{\tt z\_reio}, etc.) and pass this list to the cosmology code together with
some values. The cosmology code (in our case, {\sc class}) will read these
names and values as if they were written in an input file. If one of the names is
not understood by the cosmology code, the run stops. The advantage is that the
user can immediately write in the input file any name understood by class,
without needing to edit {\sc Monte Python}. This is not the case with {\sc
CosmoMC}. This is why users can do lots of things with {\sc Monte Python}
without ever needing to edit it or even knowing Python. If one wants to explore
a completely new cosmological model, it is enough to check that it is
implemented in {\sc class} (or to implement it oneself and recompile the class
python wrapper). But {\sc Monte Python} doesn't need to know about the change.
To be precise, in the {\sc Monte Python} input file, the user is expected to
pass the name, value, prior edge etc. of all parameters (i) to be varied; (ii)
to be fixed; (iii) to be stored in the chains as derived parameters. These can
be any {\sc class} parameter: cosmological parameters, precision parameters,
flags, input file names. Let us take two examples: 
\begin{itemize}
  \item In this paper, we showed some posterior probabilities for the angular
    diameter distance up to recombination. It turns out that this parameter is
    always computed and stored by {\sc class}, under the name {\tt `da\_rec'}.
    Hence we only needed to write in the input file of {\sc Monte Python} a
    line looking roughly like {\tt da\_rec=`derived'} (see the documentation
    for the exact syntax), and this parameter was stored in the chains. In this
    case {\sc Monte Python} did not need editing.
  \item We used in this work the parameter $[e^{-2 \tau} A_s]$. To implement
    this, there would be two possibilities. The public {\sc class} version
    understands the parameters $\tau$ and $A_s$. The first possibility is to
    modify the {\sc class} input module, teach it to check if there is an input
    parameter {\tt `exp\_m\_two\_tau\_A\_s'}, and if there is, to infer $A_s$
    from $[e^{-2 \tau} A_s]$ and $\tau$. Then there is no need to edit {\sc
    Monte Python}. However, in a case like this, it is actually much simpler to
    leave {\sc class} unchanged and to add two lines in the {\sc Monte Python}
    file  {\tt data.py}. There is a place in this file devoted to internal
    parameter redefinition. The user can add two simple lines to tell {\sc
    Monte Python} to map ({\tt `exp\_m\_two\_tau\_A\_s'}, {\tt `tau'}) to ({\tt
    `A\_s'}, {\tt `tau'}) before calling {\sc class}. This is very basic and
    does not require to know python.  All these parameter manipulations are
    particularly quick and easy with {\sc Monte Python}. 
\end{itemize}
The user is also free to rescale a parameter (e.g. $A_s$ to $10^9A_s$ in order
to avoid dealing with exponents everywhere) by specifying a rescaling factor in
the input file of {\sc Monte Python}: so this can be done without  editing
neither {\sc Monte Python} nor {\sc class}.\\
Please note however that, while this is true that any input parameter will be
understood directly by the code, to recover derived parameters, the wrapper
routine (distributed with class) should know about them. To this end, we
implemented what we think is a near-complete list of possible derived
parameters in the latest version of the wrapper.

\medskip

{\bf Playing with covariance matrices.} When chains are analyzed, the
covariance matrix is stored together with parameter names. When this matrix is
passed as input at the beginning of the new run, these names are read. The code
will then do automatically all the necessary matrix manipulation steps needed
to get all possible information from this matrix if the list of parameter has
changed: this includes parameter reordering and rescaling, getting rid of
parameters in the matrix not used in the new runs, and adding to the matrix
some diagonal elements corresponding to new parameters. All the steps are printed on
screen for the user to make sure the proper matrix is used.

\medskip

{\bf Friendly plotting.} The chains produced by {\sc Monte Python} are exactly
in the same format as those produced by {\sc CosmoMC}: the user is free to
analyze them with {\tt GetDist} or with a customized code. However {\sc Monte
Python} incorporates its own analysis module, that produce output files and one
or two dimensional plots in PDF format (including the usual "triangle plot").
Thanks to the existence of {\tt log.param} files, we just need to tell {\sc
Monte Python} to analyze a given directory - no other input is needed.
Information on the parameter best-fit, mean, minimal credible intervals,
convergence, etc., are then written in three output files with different
presentation: a text file with horizontal ordering of the parameters, a text
file with vertical ordering, and a latex file producing a latex table.  In the
plots,  the code will convert parameter names to latex format automatically (at
least in the simplest case) in order to write nice labels (e.g. it has a
routine that will automatically replace {\tt tau\_reio} by {\tt \textbackslash
tau\_$\{$reio$\}$}). If the output needs to be customized (i.e., changing the
name of the parameters, plotting only a few of them, rescaling them by some
factor, etc.), then the user can use command lines and eventually pass one
small input file with extra information. The code stores in the directory of
the run only a few PDF files (by default, only two; more if the user asks for
individual parameter plots), instead of lots of data files that would be needed
if we were relying on an external plotting software like Matlab.

\medskip

{\bf Convenient use of mock data.} The released version of {\sc Monte
Python} includes simplified likelihood codes mimicking the sensitivity of {\it
Planck}, of a {\it Euclid}-like galaxy redshift survey, and of a {\it
Euclid}-like cosmic shear survey. The users can take inspiration from these
modules to build other mock data likelihoods. They have been developed in such
way that dealing with mock data is easy and fully automatized. The first time
that a run is launched, {\sc Monte Python} will find that the mock data file
does not exist, and will create one using the fiducial model parameters passed
in input. In the next runs, the power spectra of the fiducial model will be
used as an ordinary data set. This approach is similar to the one developed in
the code FuturCMB\footnote{http://lpsc.in2p3.fr/perotto/} \cite{Perotto:2006rj}
compatible with {\sc CosmoMC}, except that the same steps needed to be
performed manually.

\section{Conclusions}

Models for the latest stages of the cosmological evolution rely on a less solid
theoretical and observational ground than the description of earlier stages,
like BBN and recombination. Reference~\cite{Vonlanthen:2010cd} suggested a way
to infer parameters from  CMB data under some assumptions about early
cosmology, but without priors on late cosmology. By standard assumption on
early cosmology, we understand essentially the standard model of recombination
in a flat Friedmann-Lema\^{i}tre universe, assuming Einstein gravity, and using
a consistency relation between the baryon and Helium abundance inferred from
standard BBN. The priors on late cosmology that we wish to avoid are models for
the acceleration of the universe at small redshift, a possible curvature
dominated stage, possible deviations from Einstein gravity on very large scale
showing up only at late times, and reionization models.

We explained how to carry such an analysis very simply, pushing the method
of~\cite{Vonlanthen:2010cd} to a higher precision level by introducing a
marginalization over the amplitude and tilt of the CMB lensing potential. We
analyzed the most recent available WMAP and SPT data in this fashion, that we
called ``agnostic'' throughout the paper. Our agnostic bounds on the minimal
``early cosmology'' model are about twice weaker than in a standard
$\Lambda$CDM analysis, but perfectly compatible with $\Lambda$CDM results:
there is no evidence that the modeling of the late-time evolution of the
background evolution, thermal history and perturbation growth in the
$\Lambda$CDM is a bad model, otherwise it would tilt the constraints on
$\omega_b$, $\omega_{cdm}$ and $n_s$ away from the ``agnostic'' results. It is
interesting that WMAP and SPT alone favor a level of CMB lensing different from
zero and compatible with $\Lambda$CDM predictions. 

We extended the analysis to two non-minimal models changing the ``early
cosmology'', with either a free density of ultra-relativistic relics, or some
massive neutrinos that could become non-relativistic before or around photon
decoupling.  In the case of free $N_{\rm eff}$, it is striking that the
``agnostic'' analysis removes any hint in favor of extra relics. The allowed
range is compatible with the standard value $N_{\rm eff}=3.046$ roughly at the
one-sigma level, with a mean smaller than three. In the case with free total
neutrino mass $M_\nu$, it is remarkable that the ``agnostic'' analysis remains
sensitive to this mass: the two-sigma bound coincides almost exactly with the
value of individual masses corresponding to a non-relativistic transition
taking place at the time of photon decoupling. 

The derivation of these robust bounds was also for us an occasion to describe
the main feature of the new parameter inference code {\sc Monte Python}, that
we release together with this paper. {\sc Monte Python} is an alternative to
{\sc CosmoMC}, interfaced with the Boltzmann code {\sc class}. It relies on the
same basic algorithm as {\sc CosmoMC}, but offers a variety of user-friendly
function, that make it suitable for a wide range of cosmological parameter
inference analyses.

\acknowledgments

We would like to thank Martin Kilbinger for useful discussions. We are also
very much indebted to Wessel Valkenburg for coming up with the most appropriate
name possible for the code. This project is supported by a research grant from
the Swiss National Science Foundation.

\end{document}